\documentclass[runningheads]{llncs}
\usepackage{graphicx}
\usepackage{color}
\usepackage[tracking=true]{microtype}
\usepackage{hyperref}
\usepackage{blindtext}

\begin{document}

\title{An Exploratory Analysis of Feedback Types Used in Online Coding Exercises}

\author{Natalie Kiesler \orcidID{0000-0002-6843-2729}
\institute{DIPF Leibniz Institute for Research and Information in Education, Frankfurt am Main, Germany, \email{kiesler@dipf.de}\\
}
}

\authorrunning{Kiesler, 2022}

\titlerunning{An Analysis of Feedback Types Used in Online Coding Exercises}

\maketitle      

\begin{abstract}
Online coding environments can help support computing students gain programming practice at their own pace. Especially informative feedback can be beneficial during such self-guided, independent study phases. This research aims at the identification of feedback types applied by CodingBat, Scratch and Blockly. Tutoring feedback as coined by Susanne Narciss along with the specification of subtypes by Keuning, Jeuring and Heeren constitute the theoretical basis. Accordingly, the five categories of elaborated feedback (knowledge about task requirements, knowledge about concepts, knowledge about mistakes, knowledge about how to proceed, and knowledge about meta-cognition) and their subtypes were utilized for the analysis of available feedback options. The study revealed difficulties in identifying clear-cut boundaries between feedback types, as the offered feedback usually integrates more than one type or subtype. Moreover, currently defined feedback types do not rigorously distinguish individualized and generic feedback. The lack of granularity is also evident in the absence of subtypes relating to the knowledge type of the task. The analysis thus has implications for the future design and investigation of applied tutoring feedback. It encourages future research on feedback types and their implementation in the context of programming exercises to define feedback types that match the demands of novice programmers.


\keywords{feedback  \and automated feedback \and feedback types \and programming education \and tutoring feedback \and online learning environments.}

\end{abstract}

\section{Introduction}
In recent years, higher education computing programs in Germany registered high numbers of student dropout~\cite{heublein2011studienabbruch,heublein2017zwischen,heubleinrichterschmelzer2020}
The corresponding studies indicate, for example, cognitively demanding tasks in the study entry phase as one of the causes for high dropouts~\cite{heublein2017zwischen,spohrer1986novice,luxton-Reilly18}.
In computing, this challenge is reflected in basic programming education, which is a common core of every study program. 
In this context, formative assessment tools providing feedback are considered a chance to help clarify the level of expected requirements, provide information and continuous practice~\cite{neugebauer2019studienabbruch,peterson2016dropout}.
\par 
Programming as core tier of Computer Science (CS) is a crucial course for first-year students. 
Fortunately, freely available learning environments, tools and plugins, such as \textit{CodingBat}~\cite{parlante2020}, \textit{CodeRunner}~\cite{moodle2020}, \textit{Alice}~\cite{pausch1995alice,cooper2000alice,garlick2010}, \textit{Blockly}~\cite{googleblockly2020,seraj2019,kiesler2016edulearn} or \textit{Scratch}~\cite{lifelongMIT,Resnick2009,wolber2011appinventor} offer numerous potentials for programming education by enabling learners to practice programming at their own pace, aiming at competency~\cite{wgreportfull2021}. 
The step-by-step support is usually implemented via hints, or providing a (partial) solution.
It is assumed that inexperienced, novel learners of programming can benefit from these interactive environments and tools, as they actively involve learners and provide additional information via feedback~\cite{chang2014effects,garlick2010,malan2007}.
\par 
Indeed, feedback is a powerful instrument to support student learning with a long and well-documented tradition~\cite{azevedo1995meta,hattie2007power,Shute2008}. 
Feedback is supposed to help identify similarities and differences between a given standard of a subject or task, or the quality of one's results. 
This is how learners' results can be improved~\cite{boud2013feedback}.
Feedback is also categorized as a crucially influencing factor on behavioral changes~\cite{epstein2002immediate,Moreno2004}.
Despite the increasing body of research on feedback, it is still challenging to fully grasp what a theoretical feedback type actually tells learners, how it should be designed, or how it can be applied in different contexts.
However, it is important for both educators and researchers to understand how feedback types manifest in practice, and to know examples of applied feedback types that are common within a context, in this case programming education. 
\par
In this work, it is assumed that novice programmers can benefit from freely available programming environments and tools, and the respective types of tutoring feedback. 
However, educators still experience challenges when implementing and designing feedback types. 
Similarly, research on feedback types and their effects is complicated by inconsistent feedback types and their application.
The research question is: 
\textit{Which types of informative tutoring feedback are applied in online coding exercises offered via CodingBat, Scratch, and Blockly?} 
The concept of informative, tutoring feedback coined by Narciss~\cite{narciss2008feedback}, as well as Keuning, Jeuring and Heeren~\cite{Keuning2016} will be used as a starting point for this analysis.
It is thus the goal to investigate the application of respective tutoring feedback types in coding exercises offered by the three popular tools to identify and describe the observed mismatch between theory and practice. 
The implications of current practices will constitute the basis for a better understanding of the design and implementation of feedback in future educational technology research and projects. 
\par 
The structure of the paper is as follows:
After this introduction, related work in form of theoretically defined feedback types and coding exercise tools will be introduced.
The research design and analysis method is outlined in section 3, before section 4 continues with the presentation of results.
The discussion of results and implications in section 5 is followed by this works' limitations, conclusions and future perspectives. 

\section{Related Work}

In this section, the term \textit{tutoring feedback} is defined and specified by subtypes. 
The presented feedback types will constitute the basis for the analysis of applied feedback types in coding exercises.
Moreover, three exercise tools for self-paced programming practice are briefly introduced: CodingBat, Scratch and Blockly.

\subsection{Definition of Tutoring Feedback}

Feedback constitutes an important factor for student learning with a long and well-documented body of research~\cite{azevedo1995meta,hattie2007power,Shute2008}. 
Boud and Molloy \cite{boud2013feedback} define feedback as a process in which learners receive information from an external source about their own work or solutions.
Feedback is supposed to help identify similarities and differences between a given standard of a subject or for any task and the quality of one's results so that learners' results can be improved~\cite{boud2013feedback}.
Thus, the term feedback describes the amount of information communicated to learners with the goal of changing their thinking, approaches or behaviors in a way that enhances learning~\cite[p.\ 154]{Shute2008}. 
As the term feedback comprises an extensive set of information, Narciss \cite[p.\ 18]{narciss2006} defines informative, \textit{tutoring feedback} as the sum of information learners receive from external sources (e.g., tutors, lecturers, compiler, automatic assessment tools, etc.) in order to cope with a given task in a context-specific situation or in a future attempt to successfully solve it.
Tutoring feedback can be distinguished from motivational feedback, summative assessment (e.g., grades) and internal feedback perceived by a subject~\cite{narciss2006,narciss2008feedback}.
\par 
In the context of programming education, the quality of information in formative settings is considered a chance for learning, as it offers several potentials with regard to individual learning paths, avoiding sources of frustration due to insufficient and cryptic compiler messages, offering advice on how to proceed and by addressing errors and error pattern.

\subsection{Types of Tutoring Feedback}
\label{sec:feedback_types}

Various forms of tutoring feedback in computer-assisted instruction scenarios are summarized by Mason and Bruning \cite[p.\ 5-6]{MasonBruning2001} via the distinction of eight levels, or with regard to complexity \cite{Shute2008}.
In this research paper, the author focuses on the five elaborated, tutoring feedback types according to Narciss \cite{narciss2006} and the specified subtypes and definitions provided by Keuning, Jeuring and Heeren \cite[p.\ 43-45]{Keuning2016}, as they match the context of programming exercises:
\begin{itemize}
    \item Knowledge about task constraints (KTC): Information related to the requirements of a task or general information how to approach a problem. 
        \begin{itemize}
            \item Hints on task requirements (TR) \item Hints on task-processing rules (TPR)
        \end{itemize}
    \item Knowledge about concepts (KC): Explanations or examples related to a concept that is addressed in a task.
        \begin{itemize}
            \item Explanations on subject matter (EXP)
            \item Examples illustrating concepts (EXA)
        \end{itemize}
    \item Knowledge about mistakes (KM): The student's error is described by a type and a level of detail.
        \begin{itemize}
            \item Test failures (TF)
            \item Compiler errors (CE)
            \item Solution errors (SE)
            \item Style issues (SI)
            \item Performance issues (PI)
        \end{itemize}
    \item Knowledge about how to proceed (KH): Information related to a student's next step, for error correction, or to come closer to a solution. 
        \begin{itemize}
            \item Bug-related hints for error correction (EC)
            \item Task-processing steps (TPS)
        \end{itemize}
    \item Knowledge about meta-cognition (KMC): Guiding or questioning students about their knowledge on knowledge.
\end{itemize}
In addition, the three rather simple, evaluative feedback types defined by Narciss  \cite{narciss2006} are recognized, even though they are generally not expected to help improve one's result or solution according to the definition of Boud and Molloy \cite{boud2013feedback}:
\begin{itemize}
    \item Knowledge of performance for a set of tasks (KP): Summative feedback indicating points or percentages of correct answers.
    \item Knowledge of result/response (KR): Information, if a student's answer is correct or incorrect.
    \item Knowledge of the correct results/response (KCR): Display of the correct answer.
\end{itemize}
The classification by Le \cite{LeNT2016} represents yet another approach towards subtypes of feedback for programming education. 
Despite the strong link to the context-specifics of programming, it seems incomplete when compared to the previously introduced specifications. 
It lacks, for instance, feedback types representing hints on meta-cognitive aspects (KMC), knowledge of the correct response (KCR), task restraint (KTC), knowledge about concepts (KC) and knowledge of performance for a set of tasks (KP).

\subsection{Tools Offering Online Coding Exercises}

Numerous approaches, prototypes and exercise tools are currently available on the web, allowing for individual practice and gaining programming experience in a self-directed manner. 
The present work aims at analyzing three popular, freely available examples in order to identify the applied types of tutoring feedback. 
Therefore, CodingBat, Scratch and Blockly are briefly introduced.

\subsubsection{CodingBat}

\begin{figure*}[htb]
  \centering
  \includegraphics[width=.99\textwidth]{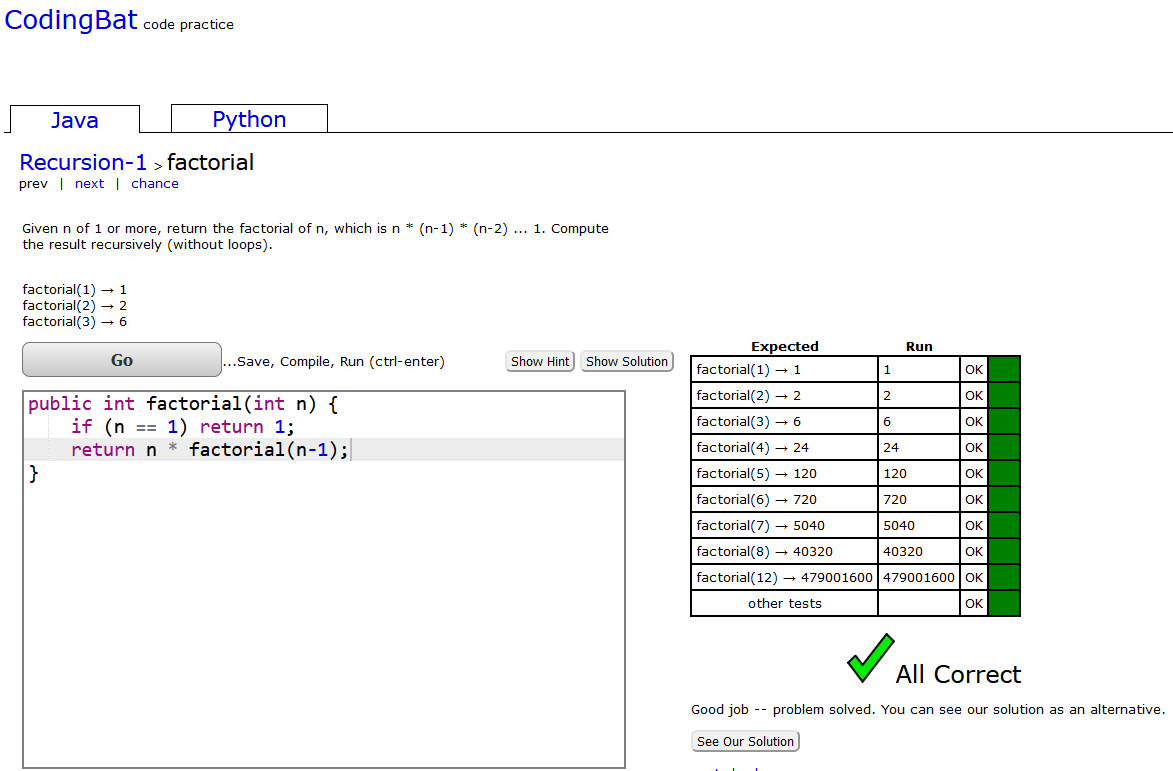}
  \caption{Screenshot of the CodingBat exercise on the faculty of n~\cite{codingbatfaculty}.}
  \label{fig:codingbat_faculty}
\end{figure*}

One of the selected tools for this analysis is CodingBat created by Nick Parlante \cite{parlante2020}, as it is freely available on the Internet.
Moreover, there is no registration or initial setup required. 
CodingBat is supposed to help train programming skills in Java and Python.
The exercises can be used individually at students' pace, but also serve as examples in face-to-face class settings, e.g., at Stanford University.
A clear advantage for users is that no preparation or installation is required to solve the tasks. 
Students can directly write a method into a white box and execute it as illustrated in \autoref{fig:codingbat_faculty}.
For some exercises, hints and solutions are available along with help pages and more worked examples.
As soon as students execute their code via the go-button, unit-tests are performed and feedback related to selected test cases is provided in the browser by means of a table.
It represents the unit-test results of the execution in terms of the expected and actual outcome related to students' input. 
A color code system (green vs.\ red) indicates, whether the unit-test results are correct and the problem has been fully solved by the student.

\subsubsection{Scratch}
\begin{figure*}[htb]
  \centering
  \includegraphics[width=.99\textwidth]{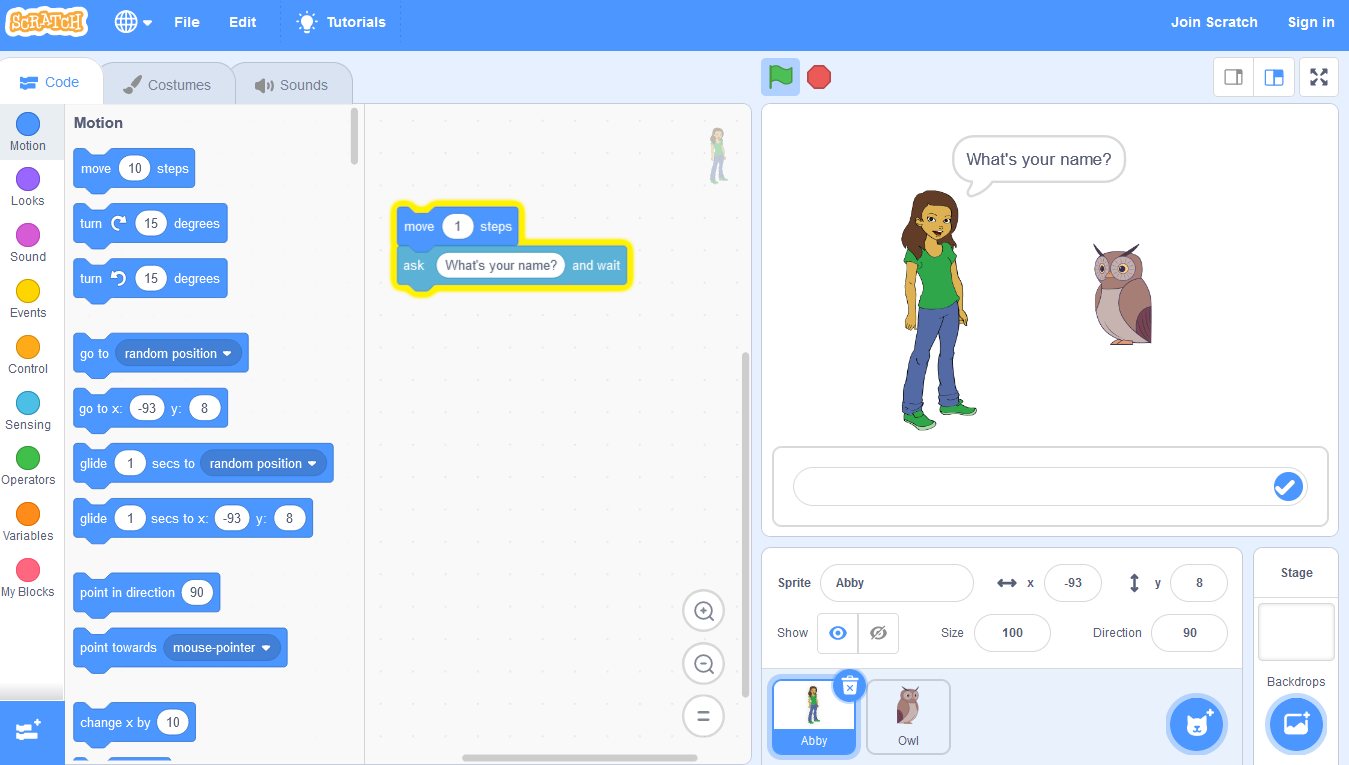}
  \caption{Screenshot of the exploratory Scratch workspace~\cite{lifelongMIT_editor}.}
  \label{fig:scratch_1}
\end{figure*}

Scratch~\cite{lifelongMIT} aims at providing a low-threshold approach towards programming.
It is a freely available visual programming language and community that allows for the simple development of stories, games and animations.
Similarly, a registration is not required to create a first project.
The potential of Scratch for formal programming education is founded the empowerment of learners to practice programming in a self-directed manner~\cite{Resnick2009}. 
Scratch utilizes visual feedback, among other types, by illustrating the execution of a constructed program and its steps in order to clarify their sequence and coherence. 
The resulting visual display (see Figure \ref{fig:scratch_1}) becomes increasingly individual, depending on the student input. 
According to Shu \cite{shu1988visual}, programming involves both hemispheres of the brain so that the use of images for learning processes in programming education is recommended.
Likewise, visual programming languages, such as Scratch, are recommended by some educators as an adequate tool for the introductory phase of programming in higher education~\cite{malan2007,tanrikulu2011users}, or at least as a supportive measure~\cite{chang2014effects}.
Today, an extensive collection of Scratch projects with stories, games and animations is available for practice.

\subsubsection{Blockly}

The Blockly open source library~\cite{googleblockly2020} allows developers to create block-based visual programs and integrate them into apps or web applications. 
The display of code blocks, the design of keywords, and how a block-based program runs, depends on the developers as well as the intended context~\cite{googleblockly2020,fraserblockly2014,pasternak2017tips}.
%
\begin{figure*}[b!]
  \centering
  \includegraphics[width=.99\textwidth]{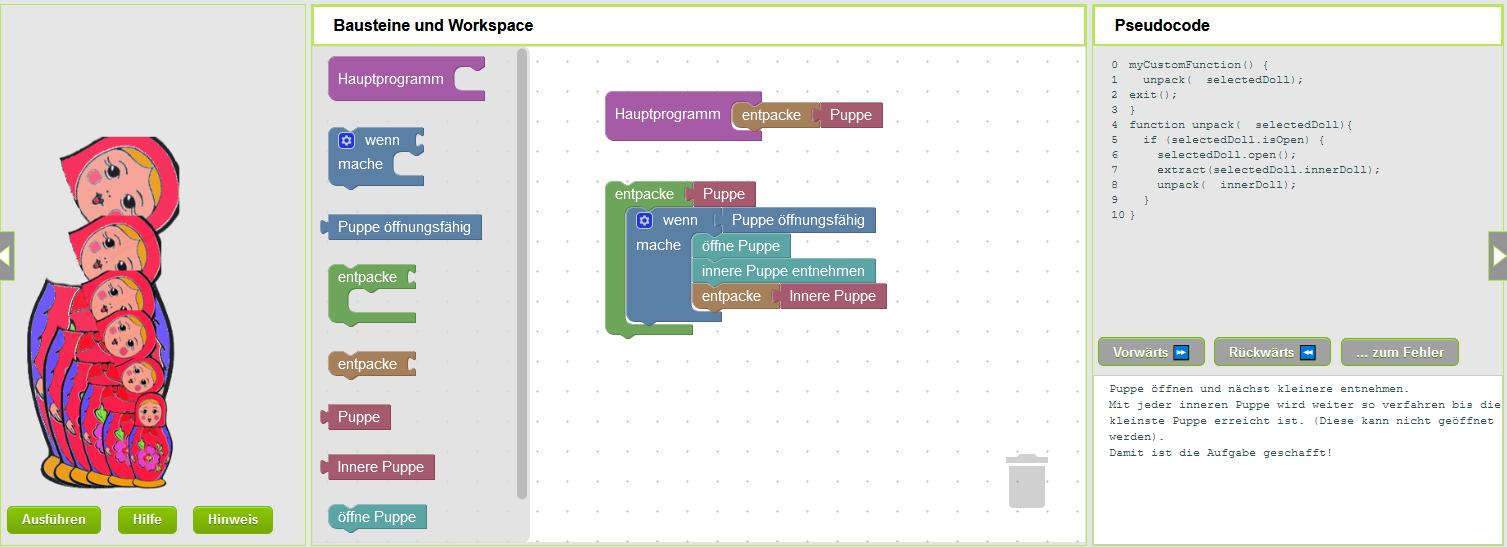}
  \caption{Screenshot of a feedback prototype that utilizes Blockly to recursively unpack a matryoshka doll~\cite{kiesler2016edulearn,kiesler2016delfi}}
  \label{fig:blockly_prototype}
\end{figure*}
Blockly is the basis for the author's development of a feedback prototype that offers a predefined selection of block-based code snippets~\cite{kiesler2016edulearn,kiesler_diss_2022}.
The implemented task asks students to develop a recursive solution for the decomposition of a closed matryoshka doll.
Students can drag and drop the ``blockly'' representations of code across the workspace and attach them to each other as illustrated in Figure \ref{fig:blockly_prototype}.
Pseudo code is generated and displayed simultaneously depending on the blocks' use and arrangement in the workspace.
As soon as learners hit the execute-button, an interactive visualizations and a short textual feedback is presented, each depending on the individual input.
Individual error messages for 168 error combinations have been prepared.
Moreover, a hint-button is available that offers additional, individual tutoring feedback.
The help-button provides support regarding the usability of the blocks and the functions of the several buttons.
A debugger enables the identification of the first incorrect block by highlighting it with a yellow frame and star.

\section{Research Design}
Despite the availability of feedback classifications, developing respective informative tutoring feedback is still not a simple task for educators.
Even though feedback types are deduced well from a number of applications~\cite[p.\ 43-45]{Keuning2016}, \cite[p.\ 19-22]{narciss2006}, and factors contributing to the information value of feedback are defined~\cite[p.\ 87]{narciss2006}, it remains challenging for educators to apply these types to their context and design informative tutoring feedback.
Moreover, the context-specific typology for programming~\cite{LeNT2016} seems to lack relevant types of feedback for fostering programming competencies.
Thus, a mismatch between the theory and the practice of tutoring feedback types used in online coding exercises is assumed. 
\par
For these reasons, it is important to identify detailed (good practice) examples of informative feedback types that match the informational needs of novice learners of programming. 
An analysis of existing, freely available, and commonly used systems can reveal how theoretical feedback types for programming can be implemented in practice. 
In the following subsections, the research question and goals will be presented, before the analysis method is introduced.


\subsection{Research Question and Goals}
The research questions of the present work is as follows: 
\textit{Which types of informative tutoring feedback are applied in online coding exercises offered via CodingBat, Scratch, and Blockly?}
The analysis of exercises that are based on these exemplary tools will lead to a better understanding of informative feedback types and how they manifest in practice. 
This understanding will help raise educators' awareness of the feedback options and potential limitations of coding tools. 
The present work will also reveal implications on the design and implementation of feedback types in the future.
Moreover, the application of the theoretical feedback types to specific tools will be a chance for future investigations of authentic feedback types' effects on students.

\subsection{Analysis}
The three well-known coding tools CodingBat, Scratch, and Blockly were selected for the analysis as they offer or support the development of informative tutoring feedback types.
As all three of them have been applied in the context of university programming education~\cite{parlante2020,malan2007,pasternak2017tips}, it is assumed that these tools will provide authentic examples of feedback that reflect on current feedback practices in the CS education community.
The initial exploration of the tools confirmed that they provide a variety of feedback types.
Their analysis will thus result in diverse examples for the feedback types and their realization, thereby answering the research question.
\par
The types of tutoring feedback defined by Narciss~\cite{narciss2006} along with the subtypes specified by Keuning, Jeuring and Heeren~\cite{Keuning2016} (see section \ref{sec:feedback_types}) constitute the analysis' theoretical basis.
One exercise of each tool was selected, respectively developed.
It should be noted that CodingBat, for example, does not offer identical feedback types for all tasks. 
Therefore, an exercise with the maximum number of feedback options was selected (see Figure~\ref{fig:codingbat_faculty}).
For Scratch, the basic editor for the creation of new projects was used~\cite{lifelongMIT_editor} (see Figure~\ref{fig:scratch_1}).
In case of the Blockly library, a prototype with an exemplary exercise was developed to support the largest possible number of feedback options (see Figure~\ref{fig:blockly_prototype}). 
Successively, implications on feedback design and research will be derived from the results of the analysis.

\section{Results}
The analysis of exercises and tools revealed the applied feedback types according to Narciss \cite{narciss2006} and the specification of subtypes by Keuning, Jeuring and Heeren~\cite{Keuning2016}. 
In the following, the identified feedback types as part of CodingBat, Scratch and Blockly exercises are presented.

\subsection{CodingBat}\label{sec:results_codingbat}
CodingBat was selected due to its freely available pool of exercises for Java and Python and, above all, its variety of feedback types. 
Even though the provided feedback does not always depend on student input, it is assumed to be helpful for learning to some extent. 
As not all types of feedback are available for every CodingBat exercise, a task with the most available feedback types was selected for the analysis:
The recursive computing of the factorial of n in Java (see Figure~\ref{fig:codingbat_faculty}).
\par 
The hint-button for this exercise leads to the following display of text, which is equal for all users:
``First, detect the ``base case'', a case so simple that the answer can be returned immediately (here when n$==$1). Otherwise make a recursive call of factorial(n-1) (towards the base case). Assume the recursive call returns a correct value, and fix that value up to make our result.''
In sum, the following feedback types are implemented:
\begin{itemize}
    \item A go-button resulting in the execution of predefined unit-tests and a color-coded table with output, which is based on students' individual code as an input (KM feedback: TF and SE; KR feedback)
    \item Compiler messages ``Compile problems'' or ``Bad code'' according to the individual input (KM feedback: CE) 
    \item A hint-button including further information independent of student input (KH feedback: TPS; KTC feedback: TR and TPR; KC feedback: EXP)
    \item A show solution-button that offers and briefly comments on one exemplary solution (KCR feedback)
    \item A help section with links to similar, worked examples and solutions (KC feedback: EXA)
\end{itemize}
It is noted that both the go-button and compiler messages are available in all CodingBat exercises. 
However, the hint-button and show solution-button are not consistently implemented for all tasks.

\subsection{Scratch}\label{sec:results_scratch}

The visual programming language scratch allows learners to literally start from scratch by providing a great variety of resources for programming, e.g., categories of blocks for motion, looks, sounds, the creation of new blocks and variables, sprites, backdrops, and much more.
Visual feedback is an important component of these multi-sensory resources and their application.
Unlike other environments or tools, scratch can be used without solving a particular problem or tasks.  
Therefore, the interface itself was used for the analysis (see Figure~\ref{fig:scratch_1}), which resulted in the identification of the following types of tutoring feedback:
\begin{itemize}
    \item A go-button in form of a green flag resulting in the visual execution of the individual arrangement of blocks (KR feedback)
    \item Tutorials and a repository with other worked examples, and projects (KC feedback: EXA)
\end{itemize}
\noindent
Unlike CodingBat, Scratch does not offer feedback related to mistakes (KM feedback).
It is an exploratory environment and test failures or compiler errors are a priori excluded by means of the block representation of the code.
The connections of each code block are thus predefined in a way that only allows syntactically valid compositions. 
Moreover, the lack of precise exercises results in a lack of task related feedback (KTC feedback).
Due to Scratch's exploratory nature, evaluative feedback types (KP and KCR feedback) are also not available. 
In sum, Scratch offers fewer feedback options than expected and as a consequence does not merge too many types. 
The strong focus on the visual mode of feedback may reduce the need for more extensive textual information and feedback.
It seems interesting to further investigate to what extent students are satisfied with these few types and the mode of the offered feedback, and whether they still have other informational needs that have not yet been met by Scratch.

\subsection{Blockly}\label{sec:results_blockly}

For the analysis of Blockly's feedback options, the open source library's developer tools were utilized~\cite{blocklydevelopertools}.
The exemplary visual display and possible connections of code blocks, keywords, and how the block-based program runs thus had to be determined by the author first (see Figure \ref{fig:blockly_prototype}).
The target group of the constructed exercise on recursion were first-year CS students, and thus novice learners of programming.
Therefore, a maximum variety of feedback types were implemented and made available to students~\cite{kiesler_diss_2022}:
\begin{itemize}
    \item Predefined connections of code blocks determining valid and invalid compositions of code, i.e., not every block can be connected to each other (KM feedback: CE and SI) 
    \item Pseudo code depending on the individual arrangement of blocks in the workspace (KC feedback: EXA)
    \item An execute-button determining the visualization of the matryoshka, the textual feedback on-screen and the additional content of the hint-button depending on the individual student input (KR feedback; KM feedback: TF; KH feedback: EC; KTC feedback: TPR; KC feedback: EXP)
    \item A hint-button including further information depending on the individual student input (KH feedback: EC and TPS ; KTC feedback: TPR; KC feedback: EXP; KMC feedback)
    \item A debugger indicating incorrect blocks by highlighting them, which depends on the individual student input (KM feedback: SE)
\end{itemize}
The prototype thus resembles the structure of other online tools and environments with programming exercises for CS students~\cite{lifelongMIT,nguyen2011alice} by providing executable code, a visualization, and hints.
The main difference, however, is that most of its feedback types are based on the individual arrangement of code blocks in the workspace so that the contents of the hint-button, for example, do not contain the same information for all users.

\section{Discussion of Implications for Feedback Design and Research}

Several indicators can be derived from the analysis of feedback types. 
These have implications on the future design and research of feedback types and their application in programming education and respective learning environments.
\par 
First of all, many feedback options of the tools and environments integrate several types of feedback into one feedback option.
For example, it is easily possible to combine KH feedback (EC and TPS), KTC feedback (TPR), KC feedback (EXP), and KMC feedback in one hint, as revealed by the CodingBat and the Blockly exercise (see \ref{sec:results_codingbat} and \ref{sec:results_blockly}).
The lack of clear-cut boundaries of feedback types in practice makes it much more difficult to investigate these types, with Scratch being the exception.
The implication for research is that it might become challenging to determine causalities, correlations, or effects of a specific feedback type. 
However, designing fewer feedback options or buttons with a one-to-one correspondence to feedback types seems unreasonable for learners in higher educational settings, as it may result in an increased cognitive load~\cite{spohrer1986novice,Paas2003,Sweller1998}.
A step-by-step analysis of students' progress when solving programming tasks and learning environments' feedback may be an approach for further research in order to identify to what extent the offered feedback matches students' informational~needs.
\par 
Moreover, the investigated feedback types do not sufficiently distinguish between general information that is provided and information that depends on learners' input. 
By definition, TPR and EXA feedback relate to general information on task constraints or concepts. 
Nonetheless, it is possible to implement these feedback types so that the information does relate to student input, as reflected in the Blockly coding exercise (see section \ref{sec:results_blockly}). 
Other feedback types, such as CE feedback with compiler error messages, are always individualized to the respective input (see section \ref{sec:results_codingbat}). 
The same is true for style issues (SI feedback).
The author therefore suggest to further distinguish between these feedback types, and add an attribute, such as $+$ or $-$ individual to the feedback~types. 
\par 
The analysis of feedback types also implies that feedback types themselves should be adapted to the context of a task, in this case programming education. 
The defined types of elaborated feedback do not yet distinguish between types of knowledge. 
Programming, however, requires different types of knowledge. 
Among them is, for example, the knowledge of elementary programming language constructs (factual knowledge), knowledge of basic characteristics of algorithms (conceptual knowledge), knowledge of basic algorithms and data structures (procedural knowledge), as well as knowledge about meta-cognition~\cite{kiesler2020iticse,kiesler2020koli}.
Therefore, it seems reasonable to specify feedback providing knowledge about concepts (KC feedback), or to add feedback types that resemble the other types of knowledge in programming.
This leads to the suggestion of two new feedback types in analogy to KC and KMC feedback:
\begin{itemize}
    \item Knowledge about facts (KF), e.g., provide key words, literals or operators 
    \item Knowledge about procedures and methods (KP), e.g., provide step-by-step guidance how to apply a method, operation or procedure by using an example
\end{itemize}
\noindent
This is how the feedback typology could resemble all four types of knowledge coined by Anderson et al.~\cite{andersonkrathwohl2001AKT}.
As programming education addresses all of these four knowledge types, implementing these additional feedback types is considered relevant for coding exercises, and therefore potentially helpful for students. 
\par
As the analyzed examples show, actual feedback of coding tools and environments can deviate from theoretical considerations or typologies.
Moreover, the multi-model representation of feedback types should receive more attention. 
Therefore, the processes of designing and investigating informative tutoring feedback both require a thorough consideration of feedback types and their implementation (not to mention timing, and other design aspects~\cite{narciss2006}). 
To conclude, an extension of the presented feedback typology to the context of programming exercises is encouraged along with further research on their implementation and effects.

\section{Limitations}

This work is limited by its specific context (e.g., programming education) and the selected coding exercises and tools. 
Results and implications cannot necessarily be transferred to any other programming environment or tool that provides informative tutoring feedback, or to other contexts.
Although the exercises themselves represent the maximum feedback options of each tool, they do not allow for generalizations in the field of computing or other related disciplines.
Moreover, the analysis excludes other feedback classifications and conceptual frameworks for high-information feedback~\cite{hattie2007power,wisniewski2020power}.

\section{Conclusions and Future Work}

Implementing informative tutoring feedback as a feature of programming exercises or tools has great potential for supporting novice learners in self-paced study phases.
In the present work, automated feedback options of exemplary CodingBat, Scratch and Blockly coding exercises were analyzed based on the feedback typology by Narciss~\cite{narciss2006} and its subtypes specified by Keuning, Jeuring and Heeren in their review of automated feedback generation for programming exercises~\cite{Keuning2016}.
The goal was to identify the applied types of feedback and to derive implications for feedback design and future research in the context of programming education.
\par
The results of the analysis reflect a number of challenges for both educators and researchers.
First of all, feedback types are often merged into one representation (e.g., in form of a prompt that appears after clicking on a hint-button). 
Thus, there is no one-to-one correspondence between theoretical and applied feedback types in practice, which complicates research on their effects. 
Second, the feedback types defined by Narciss~\cite{narciss2006}, as well as Keuning, Jeuring and Heeren~\cite{Keuning2016} do not distinguish feedback types according to their dependency on student input. 
Therefore, an attribute indicating a feedback's adaption to individual input could be useful. Third, feedback types are not distinguished in a granularity that reflects the types of knowledge required for a coding exercise (i.e., factual, conceptual, procedural and meta-cognitive knowledge). 
\par
The implications of the present work can be used as an impetus for educators and educational technology projects aiming at the design of new feedback options and messages for programming education.
The identified feedback types can further be utilized as a starting point for a qualitative investigation of feedback effects on students and their steps in the problem solving process. 
Moreover, an extended definition of feedback types for the context of programming education, perhaps based on yet another, more recent systematic literature/tool review, is encouraged along with a respective evaluation.
Certainly not all feedback types need to be adapted to learners' input or a task's knowledge type, but a more fine-grained distinction of feedback types for the context of coding exercises seems to constitute a chance for researchers, educators and learners.

\newpage
\bibliographystyle{splncs04}
\bibliography{mybibliography}
\end{document}